\documentclass[aps,prl,twocolumn,showpacs,groupedaddress]{revtex4}  
\usepackage{graphicx}  
\usepackage{dcolumn}   
\usepackage{bm}        
\usepackage{amssymb}   
\usepackage{sidecap}   

\begin{document}

\title{Interplay between couplings and common noise in phase synchronization:
disagreement between global analysis and local stability characterization}

\author{David Garc\'\i a-\'Alvarez}
\email{d.garcia-alvarez@lancaster.ac.uk}

\author{Alireza Bahraminasab}
\author{Aneta Stefanovska}
\author{Peter V.E. McClintock}

\affiliation{
\centerline{Department of Physics, Lancaster University, Lancaster LA1 4YB, United Kingdom}
}

\date{November 19, 2008}


\begin{abstract}
We consider two coupled phase oscillators in the presence of proportional (``common'') and independent white noises.
The global synchronization properties of the system are analytically studied via the Fokker-Planck equation.
When the ``effective coupling'' is big compared to the common noises, the former favors and the latter hinder synchronization.
On the contrary, when the coupling is small compared to the proportional noises, we find that the latter induce synchronization, optimally  when their intensities are big and in the  $n$:$m$ synchronization ratio.
Furthermore, in such case a small value of the coupling is better for synchronization.
Finally, we show that synchronization, which is a global property, must not be studied via local stability such as with Lyapunov exponents.

\end{abstract}

\maketitle


Synchronization of interacting oscillatory processes is ubiquitous in nature \cite{arkadybook}.
For any general system of coupled oscillators, phase dynamics approach can be usefully applied provided
that the intensities of all interactions on the oscillators, such as couplings between oscillators, effect of noise on the oscillators, etc, are small compared to the natural frequencies \cite{kuramoto}. In such case, synchronization as an adjustment of rhythms of phase coupled oscillatory processes because of the interaction with each other is well understood.
Another more intriguing kind of synchronization is the one induced by noise, which
is observed in many natural and experimental systems, such as lasers \cite{lasers}, neurons \cite{neurons}, or ecological systems \cite{ecological}.
The first theoretical approach in the study of noise-induced synchronization in phase oscillators was by analyzing the Lyapunov exponent \cite{teremae}, but Lyapunov exponent only determines the local stability. A well established method of analyzing the global behavior of the system is via the Fokker-Planck equation \cite{risken}, and synchronization in uncoupled phase oscillators with noise has already been studied using this approach \cite{goldobin, nakao}.
%

Still, an analysis that takes into account the two routes to synchronization -- interactions between the oscillators (couplings), and common noise --, and studies the interplay between each other is lacking. In fact, in most real systems, coupling among the oscillators and the action of noise are simultaneously present, e.g. \cite{kenwright, musizza}.
%
%
In this letter we systematically investigate for the first time the effect of couplings, independent noises, and  proportional noises (i.e., common noises but with different intensities) on phase synchronization. In particular, we show analytically and numerically how coupling and common noise compete in achieving
synchronization: two different routes to synchronization are possible depending on whether the couplings
or the common noise prevail. Moreover, the properties of the system when synchronized via couplings or by common noise clearly differ -- for example, whether increasing the couplings or the common noise results in stronger or weaker synchronization --, and therefore our results will be useful for taking apart systems in nature synchronized via interaction or via common noise.

The global behavior of the system will be studied via Fokker-Planck.
After reducing and solving analytically the equation, we will analyze the role of non-common noise, proportional noise, and couplings in synchronization, and we will show the competence between couplings and common noise.
Finally, we will present a case that shows that the Lyapunov exponent is not always a good indicator of synchronization.

%
%

%
%
Let us consider a system of two coupled oscillators in the presence of proportional and independent noises. The equations for the amplitude variables read:
\begin{eqnarray}
\hspace*{-1em}	\dot{x}_{{\alpha}} (t) &=&  {F}_\alpha (x_\alpha (t))
+\epsilon_\alpha\, {V}_\alpha (x_1(t), x_2(t))\nonumber\\
&&\hspace*{-4em} +\ {G}_\alpha (x_\alpha (t))\, \sqrt{D_\alpha}\, {\xi}(t)
+{H}_\alpha (x_\alpha (t))\,  \sqrt{E_\alpha}\, {\eta_\alpha}(t),\label{amplitudes}
\end{eqnarray}
for $\alpha$=1,2. Here $x_\alpha$ is the amplitude component of  the
$\alpha$-th oscillator, ${F}_\alpha$ is its individual dynamics, the ${V}$'s are the coupling functions, and the $\epsilon$'s are the coupling intensities.
${\xi}(t)$ is the common noise, and ${\eta_\alpha}(t)$ are the non-common noises. ${G}$ and ${H}$ represent the coupling of the oscillators to the noises. ${\xi}(t)$ and ${\eta_\alpha}(t)$ are assumed to be independent Gaussian white noises, with zero mean value and unit intensity (as the actual intensities were taken aside into $\sqrt{D_\alpha}$ and $\sqrt{E_\alpha}$):
 \hbox{$\langle \xi(t)\,  \xi(s) \rangle=\delta(t-s)$,}
\hbox{$\langle \eta_\alpha(t)\,  \eta_\beta(s) \rangle=\delta_{\alpha\beta}\, \delta(t-s)$,}
 \hbox{$\langle \xi(t)\,  \eta_\alpha(s) \rangle=0$.}

As long as the intensities of the couplings and of the noises remain small,
we can make phase dynamics. Applying the phase reduction method to equation (\ref{amplitudes}) we obtain the following equations for the phase variables \cite{kuramoto, nakao}:
\begin{eqnarray}
	&{ }&\dot{\phi}_\alpha (t) =  \omega_\alpha
+\epsilon_\alpha\, v_\alpha (\phi_1(t), \phi_2(t))\nonumber\\
& +&\sqrt{D_\alpha}\,  \xi(t)\, g_\alpha(\phi_\alpha(t))
+\sqrt{E_\alpha}\, \eta_\alpha(t)\, h_\alpha(\phi_\alpha(t)).\label{eqn:phases}
\end{eqnarray}
Here $\omega_\alpha$ is the natural frequency of the $\alpha$-th oscillator.
Note that, for the amplitude equations,  the noises may be either multiplicative, or just additive by making some of the functions ${G}$ and/or ${H}$ equal to 1; but the noises will become multiplicative for the phase dynamics.

We are interested in how much synchronization of a given order $n$:$m$ \cite{arkadybook} equations (\ref{eqn:phases}) yield.
We usually quantify the phase synchronization by evaluating how close to a fixed constant value the generalized phase difference $\phi_-$ stays over time. We thus make a biyective change of variables from $\phi_1$ and $\phi_2$ to $\phi_+$ and $\phi_-$:
\begin{eqnarray}
 \hspace{-2em}\phi_- = m\phi_1-n\phi_2,&
 &\phi_+ = m\phi_1+n\phi_2,\nonumber\\
\hspace{-2em}\phi_1=(\phi_+ +\phi_-)/(2m),&\quad
&\phi_2=(\phi_+ -\phi_-)/(2n).\label{eqn:phiplusminus}
\end{eqnarray}
The equation for $\phi_-$ is obtained by substituting (\ref{eqn:phases}) into the first equation of (\ref{eqn:phiplusminus}). The functions $v_\alpha$, $g_\alpha$ and $h_\alpha$ in (\ref{eqn:phases}) must be $2\pi$-periodic in all the phases, and therefore they can be written as a Fourier series. Now we will make a simplification,  by replacing such functions  by the lowest-order relevant contributions: these simplified equations will still yield all the important results. $v_1 (\phi_1(t), \phi_2(t))$ will be replaced by $a_1 \sin \phi_-$, because as long as $\phi_-$ stays close to a given constant, other harmonics correspond to fast oscillations \cite{arkadybook}. $v_2$ will be replaced by $a_2 \sin \phi_-$. The lowest resonant terms for $g_1(\phi_1)$ and $g_2(\phi_2)$ are respectively $b_1 \sin (m \phi_1)$ and $b_2 \sin (n \phi_2)$ as we will show later \cite{footnote:resonant}. Finally, the $h_\alpha$ have no resonant terms as we will show later, so we just  put the first harmonic $c_1 \sin(\phi_1)$ for $h_1$ and $c_2 \sin(\phi_2)$ for $h_2$. With only the relevant terms, equations (\ref{eqn:phases}) read
$\dot{\phi}_1  =  \omega_1
+\varepsilon_1 \sin \phi_-
 +A_1\, \xi(t)\sin (m \phi_1)
+B_1\, \eta_1(t)\sin\phi_1$, and
$\dot{\phi}_2  =  \omega_2
+\varepsilon_2\, \sin \phi_-
 +A_2\, \xi(t)\sin (n \phi_2)
+B_2\, \eta_2(t)\sin\phi_2$;
where
$\varepsilon_\alpha= a_\alpha\, \epsilon_\alpha$, $A_\alpha=b_\alpha\,\sqrt{D_\alpha}$, and $B_\alpha=c_\alpha\,\sqrt{E_\alpha}$.
Using these equations and the first equation of (\ref{eqn:phiplusminus}), we get the equation
$ \dot{\phi}_-=\omega_-+\varepsilon_-\sin\phi_-
+[m\, A_1\sin (m \phi_1)-n\, A_2 \sin (n \phi_2)]\, \xi(t)
+m\, B_1\, \eta_1(t)\sin \phi_1-n\, B_2\, \eta_2(t) \sin \phi_2$,
where $\omega_-=m\omega_1-n\omega_2$ is the detuning, and the ``effective coupling'' $\varepsilon_-$ is defined in the same way.

The former expression is an It\={o} stochastic differential equation, whose associated Fokker-Planck equation will be analyzed, considering $\phi_+$ and $\phi_-$ as the independent variables.
Things simplify much if we realise that $\phi_+(t)$ is a fast varying variable: in the absence of couplings and noises, $\phi_\alpha(t)=\omega_\alpha\, t$, so $\phi_+(t)=\omega_+\, t$. In our case with couplings and noises, some corrections have to be made to this expression; but, as long as the coupling intensities $\varepsilon_\alpha$, $A_\alpha$, $B_\alpha$ are small compared to the frequency $\omega_\alpha$, such corrections will be small compared to the quickly growing term  $\omega_+\, t$, so we can take the approximation $\phi_+(t)\approx \omega_+\, t$, and we may therefore integrate out $\phi_+$ for one period in the Fokker-Planck equation; because of that, only the terms of the Fokker-Planck equation that do not contain any derivative with respect to $\phi_+$ will be relevant. The equation for the probability density $W(\phi_+, \phi_-, t)$ is given by \cite{risken}:
\begin{eqnarray}
 \frac{\partial W}{\partial t} &=&
- \frac{\partial}{\partial \phi_-}\, [(\omega_-+\varepsilon_-\sin\phi_-)\, W]\nonumber\\
&+&\frac{1}{2}\frac{\partial^2}{\partial \phi_-^2}\{ [m\, A_1\sin (m \phi_1)-n\, A_2 \sin (n \phi_2)]^2\, W\nonumber\\
&+&[m B_1 \sin \phi_1]^2\, W
+[n B_2 \sin \phi_2]^2\, W\} +\cdots,\label{eqn:fokker}
\end{eqnarray}
where the suspension points stand for the terms that involve at least a derivative with respect to $\phi_+$. We will study the cases where the intensity of at least one noise is enough to make the system fall into the stationary solution, for which ${\partial W}/{\partial t}=0$. In that state, we can take the approximation that the probability density $W$ is almost independent of the fast variable $\phi_+$, because
as we mentioned before $\phi_+ \approx \omega_+\, t$, so $t \approx \phi_+/ \omega_+$ and therefore
$ 0={\partial W}/{\partial t}\approx
\omega_+\, {\partial W}/{\partial \phi_+}$, from which we get ${\partial W}/{\partial \phi_+}\approx 0$.
%
As a result, we can work with the probability density of $\phi_-$, $P(\phi_-)$, which results from the integration of $W$ within a period of $\phi_+$.
From the second line of (\ref{eqn:phiplusminus}), we see that the system is $4nm\pi$-periodic in the variables $\phi_+$ and $\phi_-$. $P(\phi_-)$ is then defined as
 $P(\phi_-)=({1}/{4nm\pi}) \int_0^{4nm\pi} W(\phi_+, \phi_-)d\phi_+ \approx W(\phi_-)$,
where the time dependence was taken out of $W$ because of the stationary condition, and the (approximate) independence of $W$ of $\phi_+$ was used in the last step.

From (\ref{eqn:fokker}),  replacing $W$ by $P(\phi_-)$, writing $\phi_1$ and $\phi_2$ as a function of $\phi_+$ and  $\phi_-$, and integrating out  $\phi_+$ between 0 and $4nm\pi$, we get:
%
%
\begin{eqnarray}
 &&\hspace*{-2.5em}\frac{\partial}{\partial \phi_-} \Big\{
 - (\omega_-+\varepsilon_-\sin\phi_-) P
+ \frac{1}{4}\frac{\partial}{\partial \phi_-} \big[(m^2\, (A_1^2 + B_1^2)\nonumber\\
&&\hspace*{-2.5em}+\,   n^2 (A_2^2 + B_2^2)
- 2 n m A_1 A_2  \cos\phi_-) P\big]\Big\}=\frac{\partial P}{\partial t}=0,
\label{eqn:fokkerint}
\end{eqnarray}
as the integrals of the terms that involve at least a derivative with respect to $\phi_+$ are equal to zero because of periodicity. The stationary condition was used in the last equality. The last term ``$- 2\, n\, m\, A_1\, A_2\,  \cos\phi_-$'' in the left hand side of (\ref{eqn:fokkerint}) is the contribution to the integral of the term ``$-2\, m\, A_1\sin (m \phi_1)\,n\, A_2 \sin (n \phi_2)$'' coming from the first squared bracket in (\ref{eqn:fokker}). That is the reason why we previously claimed  that the lowest-order resonant terms of $g_1(\phi_1)$ and $g_2(\phi_2)$ in (\ref{eqn:phases}) -- the couplings of the oscillators to the common noise -- were respectively $b_1 \sin (m \phi_1)$ and $b_2 \sin (n \phi_2)$: the contribution of this crossed term of the squared difference to the integral is zero except for the $m$ and $n$ harmonics respectively (and integer multiples \cite{footnote:resonant}). So other harmonics just add their ``$A^2$'' terms to the ``$B^2$'' terms arising from the non-common noises -- see $(A_1^2 + B_1^2)$ and $(A_2^2 + B_2^2)$ in (\ref{eqn:fokkerint}) --, and as a result the non-resonant harmonics of the couplings to the common noise play esentially the same role as the harmonics of the couplings to the non-common noises.

Equation (\ref{eqn:fokkerint}) reads that the expression inside the top-level bracket in the left hand side is a constant,  called the probability current $S$. If we call
 $f_1(\phi_-)=-4\, (\omega_-+\varepsilon_-\sin\phi_-)$ and
$f_2(\phi_-)= m^2\, (A_1^2 + B_1^2)  + n^2\, (A_2^2 + B_2^2)
 - 2\, n\, m\, A_1\, A_2\,  \cos\phi_-$,
equation (\ref{eqn:fokkerint}) yields
$P^{\,\prime} (\phi_-)+[f_1(\phi_-)+f_2^{\ \prime}(\phi_-)]\, P (\phi_-)/{f_2(\phi_-)}=
{S}/{f_2(\phi_-)}$.
This equation is easily solved by using the integrating factor
$M(\phi_-)=\exp[\int (f_1+f_2^{\ \prime})/{f_2}]=f_2\ \exp[\int
{f_1}/{f_2}]$,
so the general solution of the differential equation is
\begin{equation}\label{solution_raw}
 P(\phi_-)=\frac{\displaystyle S\int_{\phi_0}^{\phi_-} \exp[V(x)]\, dx\ + \ N}{\displaystyle f_2(\phi_-)\, \exp[V(\phi_-)]}\ ,
\end{equation}
where
%
 $V(x)=\int_{x_0}^x\ {f_1}/{f_2}$.
The lower limits $\phi_0$ and $x_0$ for the former integrals are arbitrary, but once such values are chosen, they have to be kept the same. For convenience, we will set both to be equal to $-\pi$. The two constants $N$ and $S$ in (\ref{solution_raw}) are obtained by imposing two conditions. The first one is that $P(\phi_-)$ has to be $2\pi$-periodic (this comes from the fact that the equation to solve (\ref{eqn:fokkerint}) is $2\pi$-periodic in $\phi_-$). The second condition is the normalization, i.e., the integral of $P(\phi_-)$ within one period has to be equal to 1.

$f_1(x)$ and $f_2(x)$ are $2\pi$-periodic functions. A primitive $\widetilde{V}(x)$ of $f_1/f_2$ can be obtained analytically, valid inside the interval $-\pi\leq x\leq \pi$. Also, the definite integral $I_V$ of $f_1/f_2$ within one period can be calculated as $\widetilde{V}(\pi)-\widetilde{V}(-\pi)$. Such integral $I_V$ turns out to be proportional to the mismatch $\omega_-$. Putting all  things together, the integral that defines $V(x)$ can always be obtained for any value of $x$, by writing  as  $x=(2\kappa-1)\pi+\varphi$, with $\kappa$ integer and $0\leq \varphi\leq 2\pi$:
$
 V(x)=\int_{-\pi}^{x} {f_1}/{f_2}=\int_{-\pi}^{(2\kappa-1)\pi} {f_1}/{f_2}+ \int_{(2\kappa-1)\pi}^{(2\kappa-1)\pi+\varphi} {f_1}/{f_2}
%
%
 =\kappa\, I_V+ \int_{-\pi}^{-\pi+\varphi} {f_1}/{f_2}
= \kappa\, I_V+\widetilde{V}(-\pi+\varphi)-\widetilde{V}(-\pi).
$
It is clear, as $I_V$ is defined as the integral within one period (or from the former expression), that
\begin{equation}\label{eqn:periodicity_V}
V(x+2\,\alpha\,\pi)=V(x)+\alpha\, I_V,\quad \alpha\in\mathbb{Z}.
\end{equation}

A relationship between the two constants $N$ and $S$ in (\ref{solution_raw}) will be obtained by imposing the periodicity condition $P(\phi_-+2\pi)=P(\phi_-)$. To write the expression for $P(\phi_-+2\pi)$, we first write the integral in (\ref{solution_raw}) as
$
 \int_{-\pi}^{\phi_-+2\pi} \exp[V(x)]\, dx=I_P+
\int_{\pi}^{\phi_-+2\pi} \exp[V(x)]\, dx
 =I_P+
\exp(I_V) \int_{-\pi}^{\phi_-} \exp[V(x)]\, dx
$,
where
$
 I_P=\int_{-\pi}^{\pi} \exp[V(x)]\ dx
$,
and (\ref{eqn:periodicity_V}) was used in the last step.
%
%
The first factor in the denominator of (\ref{solution_raw}) is periodic, $f_2(\phi_-+2\pi)=f_2(\phi_-)$. In order to write the second factor $\exp[V(\phi_-+2\pi)]$, (\ref{eqn:periodicity_V}) is used. Then, by imposing that $P(\phi_-+2\pi)$ and $P(\phi_-)$ be equal, we obtain a relationship between $S$ and $N$:
$
 S=[\exp(I_V)-1]\, N/{I_P}
$.
Note that, for $I_V$=0, or equivalently, the mismatch $\omega_-$ equal to 0, the probability current $S$ vanishes. Writing $S$ as a function of $N$ in (\ref{solution_raw}):
\begin{equation}\label{solution_onlyN}
 P(\phi_-)=\frac{N}{\displaystyle f_2(\phi_-)\, \mathrm{e}^{\, V(\phi_-)}}
 \left[1+ \frac{\mathrm{e}^{\, I_V}-1}{I_P}\int_{-\pi}^{\phi_-} \mathrm{e}^{\, V(x)}\, dx \right],
\end{equation}
where the constant $N$ is the one that makes $P(\phi_-)$ to be normalized within one period,
$
 \int_{-\pi}^{\pi} P(\phi_-)\ d\phi_- = 1
$.

\begin{center}
\begin{figure}
\includegraphics[width=0.48\textwidth]{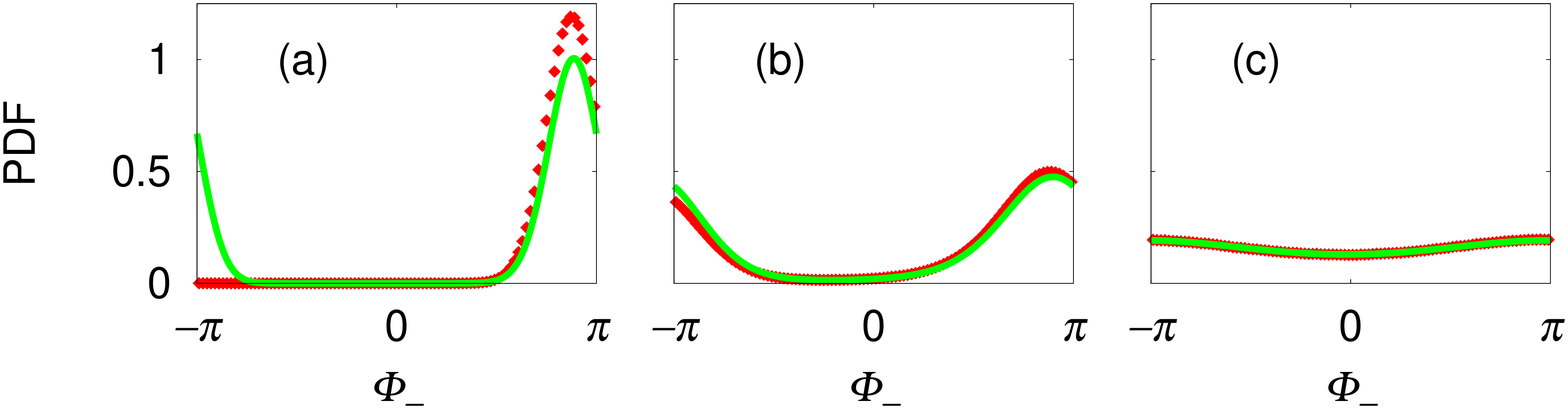}
%
%
\caption{Effect of non-common noise on synchronization. The line is the probability density function from the theoretical result (\ref{solution_onlyN}) -- the normalization constant $N$ has been numerically obtained --; the points come from simulation. The plots are for $\nu_1=1.025$, $\nu_2=0.23$, $n=9$, $m=2$, $\varepsilon_1=0.1$, $\varepsilon_2=-0.02$ -- the coupling term is $\varepsilon_{1 (2)}\sin\phi_-$ in the equation for $\dot\phi_{1 (2)}$, so $\varepsilon_1$ and $\varepsilon_2$ will have opposed signs in general --, $A_1=0.1$ and $A_2=0.02$. $B_1$ and $B_2$ are 0.1 and 0.02 in (a), 0.3 and 0.06 in (b), 1 and 0.2 in (c). From the simulation of plot (a),  $\phi_-$ stays within a small interval, the system never falls into the stationary solution due to the weakness of the noises. Therefore, for (a) the solution of the Fokker-Planck equation must be understood as an average over noise \emph{and} over initial conditions. In the same way, the probability density obtained in (a) from simulation is only valid for the initial conditions for which the simulation was performed.
\label{fig:indepnoise} }
\end{figure}
 \end{center}

\begin{center}
\begin{figure}
\includegraphics[width=0.48\textwidth]{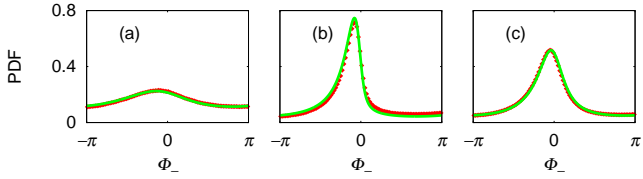}
%
\caption{Influence of the ratio between $A_1$ and $A_2$ in synchronization. $\nu_1=1.025$, $\nu_2=0.23$, $n=9$, $m=2$, $\varepsilon_1=0.001$, $\varepsilon_2=-0.0002$, $B_1=0.1$, $B_2=0.02$ and $A_2=0.1$. $A_1$ is 0.1  in (a), 0.45  in (b), 0.8  in (c).\label{fig:relatcommonnoise} }
\end{figure}
 \end{center}

Some analytical results can be derived for the case of zero frequency mismatch. Such results hold, at least approximately, as long as the  mistach remains small. For $\omega_-=0$, the expression for the probability density function becomes much simpler:
$
 P_{\, 0} (\phi_-)= N_1 [m^2(A_1^2 + B_1^2)  + n^2(A_2^2 + B_2^2)
 -\, 2\, n\, m\, A_1 A_2
 \cos\phi_-]^{(2\varepsilon_-/n\, m\, A_1 A_2)-1}
$,
where $N_1$ is a constant independent of $\phi_-$, to be adjusted so that the probability density be normalized to 1. This expression can be written in a more convenient way:
\begin{eqnarray}
 P_{\, 0} (\phi_-)&=& N_1 \biggl[m^2 B_1^2+ n^2 B_2^2 +(m\, A_1 -n\, A_2)^2 \nonumber\\
&& +\, 4\, n\, m\, A_1 A_2
 \sin^2\frac{\phi_-}{2}\biggl]^{(2\varepsilon_-/n\, m\, A_1 A_2)-1}.\qquad\label{solut_0mismatch}
\end{eqnarray}
Stronger synchronization will take place when the term depending on $\phi_-$ (the last one inside the bracket) outstands out of the $\phi_-$-independent terms (the other terms): the first three terms inside the bracket should be as close to zero as possible in order to obtain stronger synchronization. Two conclusions can be obtained from this.
In the first place, the bigger the absolute value of $B_1$ and $B_2$, the weaker the synchronization: independent noise is always harmful for synchronization. This property is plotted in Figure \ref{fig:indepnoise} (in the figures, the frequency mismatch is not exactly zero, but small). Secondly, the third term will be closer to zero the smaller in absolute value $m\, A_1 -n\, A_2$ be. In other words: the values of the couplings of the proportional noise to the first and second oscillators should ideally be in the same $n$:$m$ synchronization ratio in order to enhace the strenght of synchronization. This property can be seen on Figure \ref{fig:relatcommonnoise}: for (b), $m\, A_1 -n\, A_2$ is equal to 0, and the corresponding probability density is the most localized. Because of this, we will assume from now on that $A_1$ and $A_2$ both have the same sign.
%
\begin{center}
\begin{figure}
\includegraphics[width=0.48\textwidth]{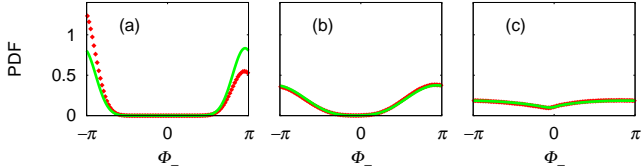}
 %
\caption{Influence of the intensities of the common noise in coupling-induced synchronization. $\nu_1=1.025$, $\nu_2=0.23$, $n=9$, $m=2$, $\varepsilon_1=0.3$, $\varepsilon_2=-0.05$, $B_1=0.1$, $B_2=0.02$.  $A_1$ and $A_2$ are 0.225 and 0.05 in (a); 0.45 and 0.1 in (b); and 0.675 and 0.15 in (c) -- note that $m\, A_1 -n\, A_2=0$ in the three plots --. Here we have for (a) the same as in Figure \ref{fig:indepnoise}(a).
\label{fig:coupling_induced} }
\end{figure}
 \end{center}
\begin{center}
\begin{figure}
\includegraphics[width=0.48\textwidth]{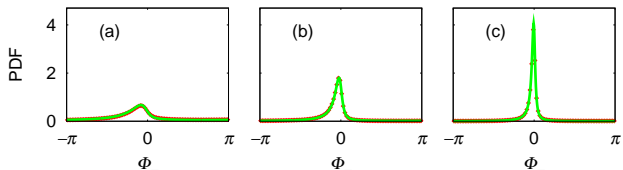}
%
\caption{Synchronization increases with the intensities of the common noise when induced by the latter instead of by couplings. $\nu_1=1.025$, $\nu_2=0.23$, $n=9$, $m=2$, $\varepsilon_1=0.01$, $\varepsilon_2=-0.002$, $B_1=0.1$, $B_2=0.02$.  $A_1$ and $A_2$ are 0.45 and 0.1 in (a); 0.9 and 0.2 in (b); 1.8 and 0.4 in (c) -- note that $m\, A_1 -n\, A_2=0$ in the three plots --.
\label{fig:noise_induced} }
\end{figure}
 \end{center}

\vspace{-6em}

After stating the former constraint for $A_1$ and $A_2$ in order to improve synchronization, we now study how their intensities affect. Here we have two different scenarios: from the exponent in (\ref{solut_0mismatch}) we can study the competence between coupling and common-noise in synchronization. When the ``effective coupling'' $\varepsilon_-$ is big compared to $A_1\, A_2$, more specifically $\varepsilon_- > n\, m\, A_1\, A_2/2$, the exponent is positive, so the probability density will have its maximum at $\phi_-=\pi$. If the product $A_1\, A_2$ increases, the bracket in (\ref{solut_0mismatch}) will depend more strongly on $\phi_-$ -- therefore the contribution of the bracket will be towards improving synchronization -- but, on the other hand, the exponent  will be smaller -- therefore the contribution of the exponent will be towards flattening the probability density, and therefore towards hindering synchronization --. As (\ref{solut_0mismatch}) is more sensitive to the second effect (the exponent) than to the first effect, we conclude that,  in this case of \emph{coupling-induced synchronization}, bigger values of the noise (either common or non-common) hinder synchronization. This property can be seen on Figure \ref{fig:coupling_induced}.

When the intensities $A_1$ and $A_2$ of the common noise are big enough compared to the effective coupling, more specifically when $\varepsilon_- < n\, m\, A_1\, A_2/2$, we are in a different regime: \emph{common-noise-induced synchronization}.
 As the  exponent in (\ref{solut_0mismatch}) is negative, the probability density will have a maximum at $\phi_-=0$. Different from the previous case, a big value of $A_1\, A_2$ is good now in the two ways: it makes the  $\phi_-$-dependent term inside the bracket outstand out of the other terms, and it makes the exponent be bigger in absolute value. This is shown on Figure \ref{fig:noise_induced} (and by comparing Figure \ref{fig:relatcommonnoise}(c) with \ref{fig:relatcommonnoise}(a)). Therefore, we can have very narrow peaks in the probability density and, as a result, strong synchronization, for small values of the non-commmon noises, and  big values of the couplings of the common noise to the oscillators, whose ``generalized difference'' $m\, A_1 -n\, A_2$ is small in absolute value (nevertheless, $A_1$ or $A_2$ cannot be arbitrarily big because the approximation $\phi_+\approx\omega_+\, t$ would not hold any more; furthermore, from the physical standpoint, we must recall that the phase dynamics approach is valid when the values of the couplings are not big). In this regime of common-noise-induced synchronization, a bigger value of the effective coupling $\varepsilon_-$ will result into an exponent smaller in absolute value: \emph{small} values for the couplings between the oscillators are better for common-noise-induced synchronization.


Finally, we show that local stability, usually studied by means of the Lyapunov exponents, does not always caracterize the synchronization properties of the system. In this letter we will just present a counter-example, and a more thorough analysis will be published elsewhere. Let us consider the most favorable case for common-noise-induced synchronization, where $B_1=B_2=m\, A_1 -n\, A_2=\omega_-=0$. As we have seen before, the probability density will have a maximum at $\phi_-=0$. Equation (\ref{eqn:fokkerint}) becomes
 $\partial_{\phi_-}\, (
 -\, \varepsilon_-\sin\phi_-\, P)
+\partial_{\phi_-}^{\ 2} [
 \, n\, m\, A_1\, A_2\,  \sin^2 (\phi_-/2)\, P]=\partial_t P
$.
The corresponding It\={o} stochastic differential equation is:
$
d\phi_-=\varepsilon_-\sin\phi_-\, dt+\sqrt{2\, n\, m\, A_1\, A_2\,  \sin^2 (\phi_-/2)}\ \, dw
$,
where $w$ is a Wiener process \cite{risken}. By using It\={o}'s lemma \cite{itolemma}, we get:
$
\langle d\log|\phi_-|\rangle=(\varepsilon_-\sin\phi_-/\phi_- -\, n\, m\, A_1\, A_2\,  \sin^2 (\phi_-/2)/\phi_-^{\, 2})\, dt
$.
By taking the limit $\phi_-\to 0$, we get the Lyapunov exponent at $\phi_-=0$:
\begin{equation}\label{lyapunov}
\lambda= \varepsilon_- -\frac{n\, m\, A_1\, A_2}{4}.
\end{equation}
Therefore, in the case $n\, m\, A_1\, A_2/4<\varepsilon_-<n\, m\, A_1\, A_2/2$, the global analysis we have made via the probability density yields that the common noise prevails over the couplings in inducing synchronization, and that the system synchronizes with $\phi_-$ staying close to 0 most of the time. Nevertheless, the Lyapunov exponent is positive at $\phi_-=0$. This shows that the local analysis made via the Lyapunov exponent sometimes fails when characterizing synchronization, which is a global property.

In conclusion, we have studied phase synchronization between two oscillators in the general case: with couplings between each other, and the action of proportional and independent noises.
Two different routes to synchronization are possible: via couplings and via common noise. When one induces synchronization, the other hinders it, thereby showing the competence of the two ways of synchronization. In particular, we have stated a surprising property: when synchronization is achieved via common noise, smaller values of the couplings (ideally no couplings) favor it.
 Finally, we have shown that synchronization, that is a global property, must not be studied by means of a local stability analysis such as via Lyapunov exponents. This work was supported by EU Project BRACCIA.


\begin{thebibliography}{99}
\bibitem{footnote:resonant} We assume that $n$ and $m$ have no common divisors, otherwise we divide both by the greatest common divisor. In the general case, all terms $\sin (k \phi_-)$ for integer $k$ are resonant in $v$. Likewise, all the terms $\sin (k m \phi_1)$ are relevant in $g_1$ and all the terms $\sin (k n \phi_2)$ are relevant in $g_2$, although resonance only takes place between terms with the same $k$.
\end{thebibliography}
\end{document}